\documentstyle[twoside,fleqn,espcrc2,epsf]{article}

\def\lapx{\,\,\lower 2pt \hbox{$\buildrel<\over{\scriptstyle{\sim}}$}\,\,}

\newcommand{\AmS}{{\protect\the\textfont2
  A\kern-.1667em\lower.5ex\hbox{M}\kern-.125emS}}

\title{S-wave charmed mesons in lattice NRQCD}

\author{Randy Lewis\address{Jefferson Lab, 12000 Jefferson Avenue,
        Newport News, VA, U.S.A. 23606 \\ and Department of Physics,
        University of Regina, Regina, SK, Canada S4S 0A2}
        and 
        R. M. Woloshyn\address{TRIUMF, 4004 Wesbrook Mall, Vancouver, BC,
        Canada V6T 2A3}}
       
\begin{document}

\begin{abstract}
Heavy-light mesons can be studied using the $1/M$ expansion of NRQCD,
provided the heavy quark mass is sufficiently large.  Calculations of the 
S-wave charmed meson masses from a classically and tadpole-improved
action are presented.  A comparison of $O(1/M)$, $O(1/M^2)$ and
$O(1/M^3)$ results allows convergence of the expansion to be discussed.
It is shown that the form of discretized heavy quark propagation must
be chosen carefully.
\end{abstract}

\maketitle

\section{INTRODUCTION}

Nonperturbative strong dynamics is typified by the mass scale 
$\Lambda_{\rm QCD}$.  Interactions involving a very heavy quark of mass $M$
can be studied systematically by expanding in $\Lambda_{\rm QCD}/M$.
Upon truncation of the expansion at some order, the resulting effective 
theory is not renormalizable and requires a momentum cutoff.

If the required regularization is performed via a space-time lattice, then
the cutoff is proportional to the inverse lattice spacing $1/a$.  A useful
effective theory must satisfy
\begin{equation}
   \Lambda_{\rm QCD} \ll 1/a \lapx M,
\end{equation}
so the cutoff is large enough to include the bulk of the nonperturbative
dynamics in the low-energy effective theory, but small enough that the
truncation of the expansion remains sensible.

Is the charm quark heavy enough for a useful lattice effective theory?
The present work addresses this question through a study of the masses
of S-waves charmed mesons using quenched lattice NRQCD.\cite{NRQCD}  
Calculations are performed at two lattice
spacings near 0.22 fm and 0.26 fm, and in each case results are given
separately at $O(1/M)$, $O(1/M^2)$ and $O(1/M^3)$ in the effective
theory.  This work is an extension of results that have been reported
previously\cite{LewWol}, and further details can be found in that paper.
Other authors have considered NRQCD up to $O(1/M^2)$\cite{others}.

\section{ACTION}

The lattice action has three terms: gauge action, light quark action
and heavy quark action.  
The entire action is classically and tadpole-improved with
the tadpole factor defined by
\begin{equation}  
   U_0 = \left<\frac{1}{3}{\rm ReTr}U_{\rm pl}\right>^{1/4}.
\end{equation}
The gauge action includes a sum over 1$\times$2 rectangular
plaquettes as well as 1$\times$1 elementary plaquettes.  
For light fermions, the Sheikholeslami-Wohlert action\cite{SW} is used
with the clover coefficient set to its tadpole-improved value.
The heavy quark action is NRQCD.

A discretization of the NRQCD action leads to the following
Green's function propagation\cite{NRQCD}:
\begin{eqnarray}
G_1\!\!\! &=& \!\!\!\left(1-\frac{aH_0}{2n}\right)^n\frac{U_4^\dagger}{U_0}
        \left(1-\frac{aH_0}{2n}\right)^n\,\delta_{\vec{x},0}, \\
G_{\tau+1}\!\!\! &=&\!\!\!
       \left(1-\frac{aH_0}{2n}\right)^n\frac{U_4^\dagger}{U_0}
    \left(1-\frac{aH_0}{2n}\right)^n \nonumber \\
    \!\!\! && \!\!\! \times (1-a\delta{H})G_\tau~~,~~\tau>0, \label{G2}
\end{eqnarray}
where ``$n$'' should be chosen to stabilize the numerics, and the
Hamiltonian is
\begin{eqnarray}
H \!\!\! &=& \!\!\! H_0 + \delta{H}, \\
\delta{H}\!\!\! &=& \!\!\!
        \delta{H}^{(1)} + \delta{H}^{(2)} + \delta{H}^{(3)} + O(1/M^4),\\
H_0 \!\!\! &=& \!\!\! \frac{-\Delta^{(2)}}{2M}, \\
\delta{H}^{(1)} \!\!\! &=& \!\!\! 
                 -\frac{c_4}{U_0^4}\frac{g}{2M}\mbox{{\boldmath$\sigma$}}
                    \cdot\tilde{\bf B} + c_5\frac{a^2\Delta^{(4)}}{24M}, \\
\delta{H}^{(2)}\!\!\! &=& \!\!\!
                   \frac{c_2}{U_0^4}\frac{ig}{8M^2}(\tilde{\bf \Delta}\cdot
                    \tilde{\bf E}-\tilde{\bf E}\cdot\tilde{\bf \Delta})
                   - c_6\frac{a(\Delta^{(2)})^2}{16nM^2} \nonumber \\
   \!\!\! &&\!\!\!  -\frac{c_3}{U_0^4}\frac{g}{8M^2}\mbox{{\boldmath$\sigma$}}
               \cdot(\tilde{\bf \Delta}\times\tilde{\bf E}-\tilde{\bf E}\times
                 \tilde{\bf \Delta}), \\
\delta{H}^{(3)}\!\!\! &=&\!\!\! -c_1\frac{(\Delta^{(2)})^2}{8M^3}
                    -\frac{c_7}{U_0^4}\frac{g}{8M^3}\left\{\tilde\Delta^{(2)},
                        \mbox{{\boldmath$\sigma$}}\cdot\tilde{\bf B}\right\}
                    \nonumber \\
               \!\!\!  && \!\!\! -\frac{c_9}{U_0^8}\frac{ig^2}{8M^3}
                        \mbox{{\boldmath$\sigma$}}\cdot
                        (\tilde{\bf E}\times\tilde{\bf E}
                        +\tilde{\bf B}\times\tilde{\bf B})
                    \nonumber \\
         \!\!\! && \!\!\!-\frac{c_{10}}{U_0^8}\frac{g^2}{8M^3}(\tilde{\bf E}^2
                    +\tilde{\bf B}^2) 
                    -c_{11}\frac{a^2(\Delta^{(2)})^3}{192n^2M^3}
                    \nonumber \\
        \!\!\! && \!\!\! + {\rm purely~quantum~effects}, \label{H3} \\
\tilde{E}_i \!\!\! &=& \!\!\! \tilde{F}_{4i}, \\
\tilde{B}_i \!\!\! &=& \!\!\! \frac{1}{2}\epsilon_{ijk}\tilde{F}_{jk}.
\end{eqnarray}
A tilde denotes removal of the leading discretization errors.
Classically, the coefficients $c_i$ are all unity, and their nonclassical
corrections will not be discussed in this work.

It should be noted that 
the separation of $H_0$ and $\delta{H}$ in Eq.~(\ref{G2}) is not unique.
The heavy quark propagation of Eq.~(\ref{G2}) uses a simple linear
approximation to the true exponential dependence on $\delta{H}$, while
using a better-than-linear approximation for $H_0$.
The present work will report on a generalization of this choice.

\section{RESULTS}

All data presented here correspond to a charmed meson with a light quark 
mass that is roughly twice the strange quark mass.  More extensive results,
including bottom mesons, can be found in Ref.~\cite{LewWol}.
The data sample includes 400(300) gauge field configurations at $\beta=
6.8(7.0)$ corresponding to $a \approx 0.26$fm(0.22fm).  
All plots include bootstrap errors from 1000 ensembles.

\begin{figure}[hb]
\epsfxsize=180pt \epsfbox[30 490 498 710]{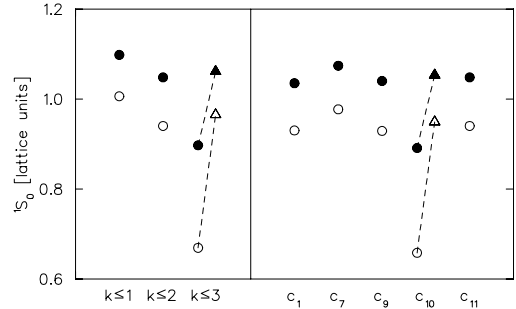}
\caption{The simulation energy of a ground state charmed meson at rest,
         up to $O(1/M^k)$.
         Solid(open) symbols denote data at $\beta=6.8(7.0)$.
         Triangles are produced by subtracting the vacuum expectation 
         value from the $c_{10}$ term.
         To the right of the vertical line, the effect of adding
         each $O(1/M^3)$ term to the $O(1/M^2)$ Hamiltonian is shown
         individually.
        }\label{Eord3}
\end{figure}
Fig.~\ref{Eord3} shows the simulation energy (as read from the plateau
of an effective mass plot) of the ${}^1S_0$ meson.  Notice that at both
lattice spacings the
$O(1/M^3)$ contribution is twice as large as the $O(1/M^2)$ contribution,
and that this large effect is dominated by the term containing $c_{10}$ in
the Hamiltonian, Eq.~(\ref{H3}).
This term is unique because it is the only term up to $O(1/M^3)$ which
contains a nonzero vacuum expectation value.  The vacuum value can be
calculated from our gauge field configurations,
and as shown in Fig.~\ref{Eord3} the simulation energy displays a very
pleasing $1/M$ expansion after removal of the vacuum value.

\begin{figure}[hb]
\epsfxsize=180pt \epsfbox[30 490 498 710]{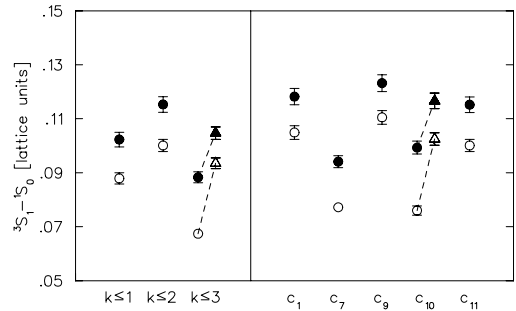}
\caption{The spin splitting of S-wave charmed mesons, with notation
         as in Fig.~\protect\ref{Eord3}.
        }\label{hypord3}
\end{figure}
The mass difference between ${}^3S_1$ and ${}^1S_0$ mesons is shown 
in Fig.~\ref{hypord3}.  When the vacuum expectation value is not
removed from the Hamiltonian, the $O(1/M^3)$ contribution is
twice as large as the $O(1/M^2)$ contribution in magnitude, 
due to large effects from the terms containing $c_7$ and $c_{10}$.
Fig.~\ref{hypord3} indicates that the $c_{10}$ effect is entirely due
to the vacuum expectation value.

This very substantial dependence of the spin splitting on the vacuum value
should be disturbing, since the $c_{10}$ term is spin-independent.
Apparently the vacuum value is so large that it destabilizes
heavy quark propagation and thereby introduces a spurious effect
unless the vacuum value is removed from the Hamiltonian.
To support this claim, we have redone the calculation after subtracting
the $c_{10}$ term from $\delta{H}$ and adding it to $H_0$ in Eq.~(\ref{G2}).
When this is done, the triangles of Fig.~\ref{hypord3} are reproduced 
regardless of whether the vacuum value is subtracted from the Hamiltonian.  
This is the physically-expected result.

Still, Fig.~\ref{hypord3} contains a large $O(1/M^3)$ contribution
dominated by the term containing $c_7$.
Although this term does not contain a nonzero vacuum expectation value,
one wonders if the heavy quark propagation might be unstable for this term
as well, unless a better-than-linear approximation is used for the $c_7$
term in Eq.~(\ref{G2}).

\begin{figure}[hb]
\epsfxsize=180pt \epsfbox[30 490 498 710]{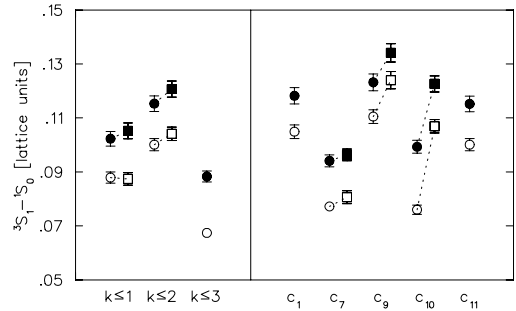}
\caption{The data (circles) of Fig.~\protect\ref{hypord3}, which use
         Eq.~(\protect\ref{G2}), are compared to
         data (squares) obtained from Eq.~(\protect\ref{G2new}).
         The vacuum expectation value is not removed.
        }\label{newfig}
\end{figure}

Fig.~\ref{newfig} shows the effect of subtracting all of $\delta{H}$ 
from its present location in Eq.~(\ref{G2}) and putting the full Hamiltonian
in place of $H_0$.
That is,
\begin{eqnarray}
G_{\tau+1} \!\!\!&=&\!\!\! \left(1-\frac{aH}{2n}\right)^n\frac{U_4^\dagger}{U_0}
    \left(1-\frac{aH}{2n}\right)^n G_\tau, \label{G2new}
\end{eqnarray}
and we choose $G_0 \equiv \delta_{\vec{x},0}$.
To maintain classical improvement, the following $O(1/M^3)$ term must be
added to the Hamiltonian:
\begin{equation}
   \delta{H}_{\rm new} = -\frac{a}{4n}\left\{
                    \left(H_0+\delta{H}^{(1)}\right),\delta{H}^{(2)}\right\}.
\end{equation}
Simulations with $n=5$ and $n=7$ are indistinguishable.
At both lattice spacings, the large $c_7$ effect is found to be robust,
and the contribution from $c_9$ tends to increase.
No discussion of the spin-independent terms containing $c_1$ and $c_{11}$ 
is presented here.

\section{DISCUSSION}

Instabilities can arise in the NRQCD expansion for charmed mesons 
due to the presence of a large vacuum expectation value.
A better-than-linear approximation to heavy quark propagation is
valuable for ensuring stability.

Substantial effects on spin splitting were found from $O(1/M^3)$
terms in the action.  However further study, for example, of alternative
definitions for the tadpole factor or of perturbative improvement
for the NRQCD coefficients is needed before one can reach a definitive
conclusion about the convergence of the NRQCD expansion for charmed mesons.

\section*{ACKNOWLEDGMENTS}

The authors thank G. P. Lepage, N. Shakespeare and H. Trottier for discussions.
This work was supported in part by the Natural Sciences and Engineering
Research Council of Canada, and the U.S.
Department of Energy, contract DE-AC05-84ER40150.

\end{document}